\documentclass[12pt]{article}
\usepackage{graphicx}
\begin{document}
\thispagestyle{headings}
\begin{center}
\large \bf Extracellular-to-intracellular signal transfer via 
G-proteins 

\vspace{.5cm}\footnotesize 

Gabriele Scheler, Department of Radiology,
Stanford University {\tt scheler@stanford.edu}
\end{center}

\small

\begin{abstract}

We look at the problem of signal transduction by 
extracellular agonist binding to a receptor protein at the 
membrane (sensor) via binding of G-proteins (effectors) 
to a highly integrative target molecule, such as the 
second messenger cAMP (target).
We explore the effects of  
binding times, effector assignment and effector pool size on the 
shape of the output signal under different input scenarios.
We conclude that low rates of information transfer may sometimes 
coincide with a high probability or efficiency of plasticity induction.

\end{abstract}

\noindent{\bf Keywords:} signal integration, G-protein, 
intracellular signaling, cAMP, calcium, plasticity

\vspace{.5cm}
{\large Introduction.} 

\begin{figure}[htb]
\begin{center}
\includegraphics[width=8cm]{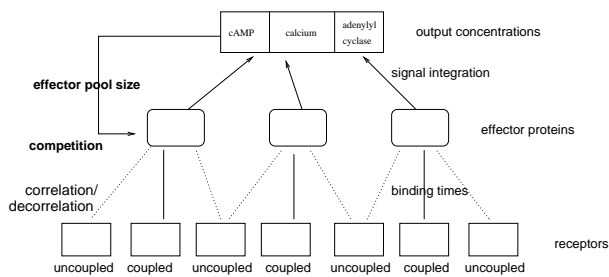}
\caption{\footnotesize Competition for effectors and feedback regulation of effector pool size}
\label{figure-1}
\end{center}
\end{figure}

G-protein coupled signal transduction proceeds by receptor activation from 
extracellular ligands and intracellular transduction through a potentially 
limited and dynamically regulated pool of effector molecules (G-proteins, 
such as $G_s$ proteins, $G_i$ proteins). Evidence for 
this is found in the observation that certain receptors (e.g. dopamine D2 
receptors) exist in distinct affinity states (low affinity vs. 
high affinity for the native ligand) dependent on their coupling to an effector 
G-protein \cite{SeemanPetal2005}.
We suggest a model with a set of input units for a number of {\it receptors},
a second set of units for {\it effectors} which can be linked to the receptors,
and an output set of units for regulated {\it concentrations} as target values 
(e.g. adenylyl cyclase, calcium, cAMP). The binding times between receptors 
and effectors are regulated by proteins such as GRkinases, RGS or calcium 
sensors (NCS-1) \cite{GainetdinovRRetal2004}.
The number of effectors is typically lower than the number of receptors,
such that a number of receptors are 'running empty', i.e. they do not 
transmit any signals they receive to the output (see Fig.~\ref{figure-1}).
Vice versa, overexpression of the active $G_{\beta\gamma}$ components has been 
shown to increase signal transduction.

The system operates by 
signal transmission from receptor to effector via variable binding times, 
a variable assignment of effectors to receptors via local competition 
(on a slower time-scale), and dynamical adjustment of 
effector pool size 
controlled by temporal integration over multiple output 
values in an encompassing feedback loop
(see Fig.~\ref{figure-1}).

\vspace{.5cm}
{\large Regulation of Binding Times.}

At the level of receptor-G-protein 
interaction, regulation of GR kinases and RGS proteins 
plays a decisive role in the rate of binding and signaling of effector 
proteins to receptors.
A critical issue for membrane receptors is temporal integration
of signals from different ligands which have effects on the same target
\cite{Scheler2004}.

A NxJ matrix of inputs (with N sensors and a time series of inputs of length 
J) becomes transformed into an output vector of length J by mapping 
through a MxJ matrix, where M is dynamically regulated but typically smaller 
than N ({\it filtering}). Each sensor is linked to only one effector at a time.
When a sensor receives a signal, it binds to an effector, which transmits 
a quantal effect on the target concentration, the amplitude of 
which may be additionally modulated (e.g., by the effects 
of adenylyl cyclases).

\begin{figure}[htb]
\begin{center}
\includegraphics[width=8cm]{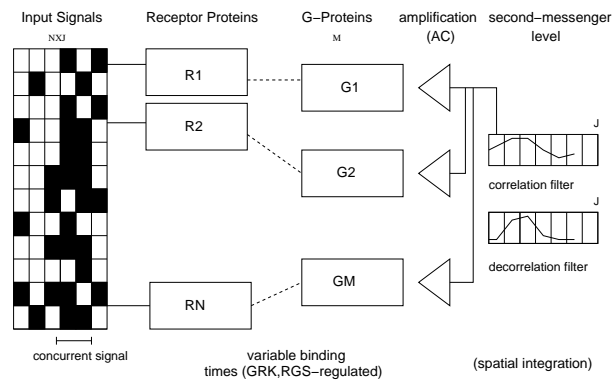}
\caption{\footnotesize Shapes of the output signal regulated by 
receptor-effector binding times. Input signals with periods of 
high or low concurrency are modulated by a cross-correlation filter 
defined by receptor-effector binding times. If binding times are long,  
correlation is increased (=correlation filter), if they are short, it 
is decreased (=decorrelation filter)}
\label{signal-transduction}
\end{center}
\end{figure}

Now for several values of M, and different input signals, 
we can produce a raw output vector based on a summed integration of the
effector values  
(see Fig.~\ref{signal-transduction}).

The dynamical properties of second messenger concentration 
are an important part of their function, as has been shown 
for calcium. Brief, phasic increases 
have a different effect on downstream signaling than 
longer-lasting broader signals. 
By altering the binding times of R-G coupling 
\cite{GainetdinovRRetal2004}, input correlation can thus be increased 
or decreased by a cell-internal cross-correlation filter.
 
\begin{figure}[htb]
\begin{center}
\includegraphics[width=6cm]{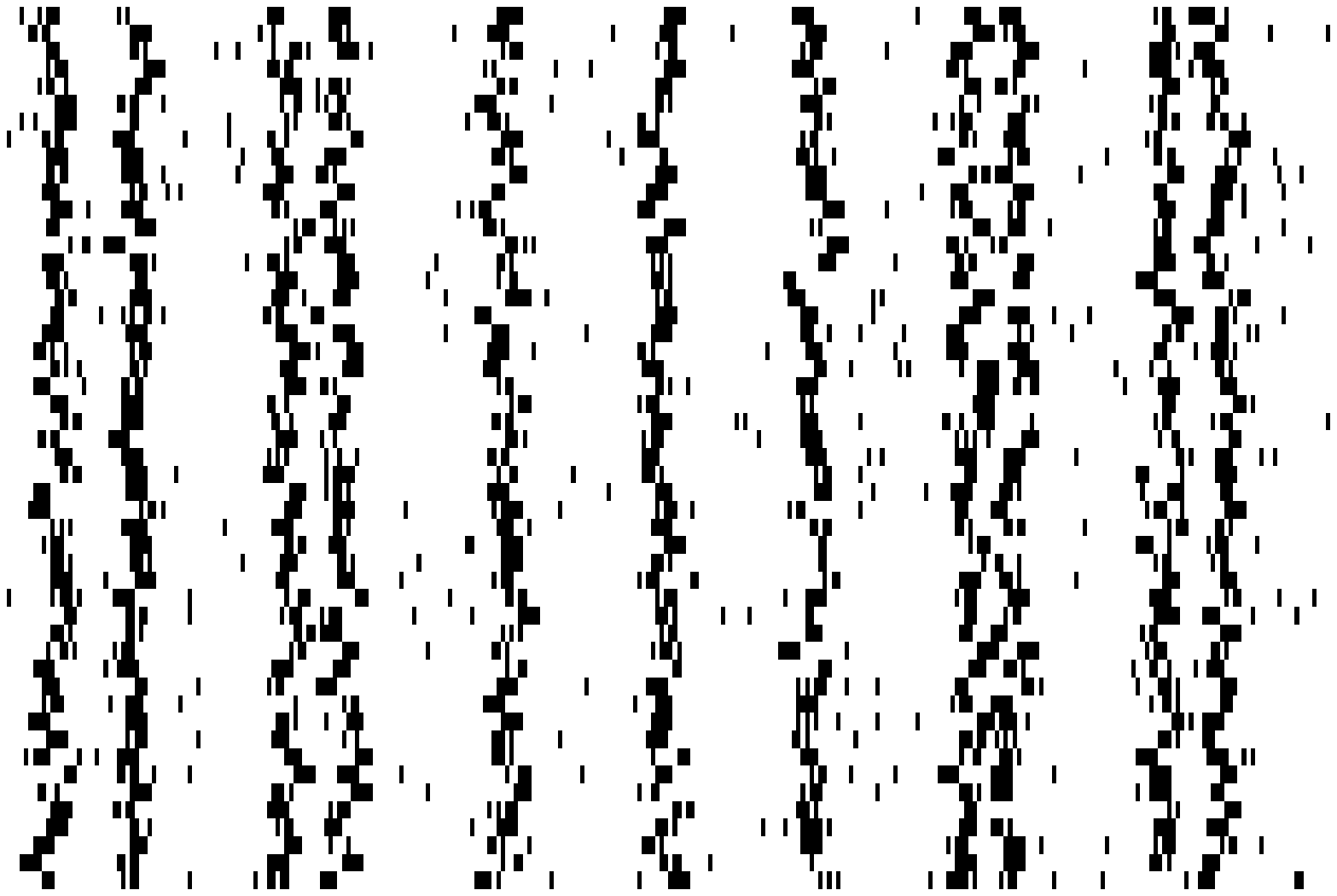}
\hspace*{0.5cm}
\includegraphics[width=5cm]{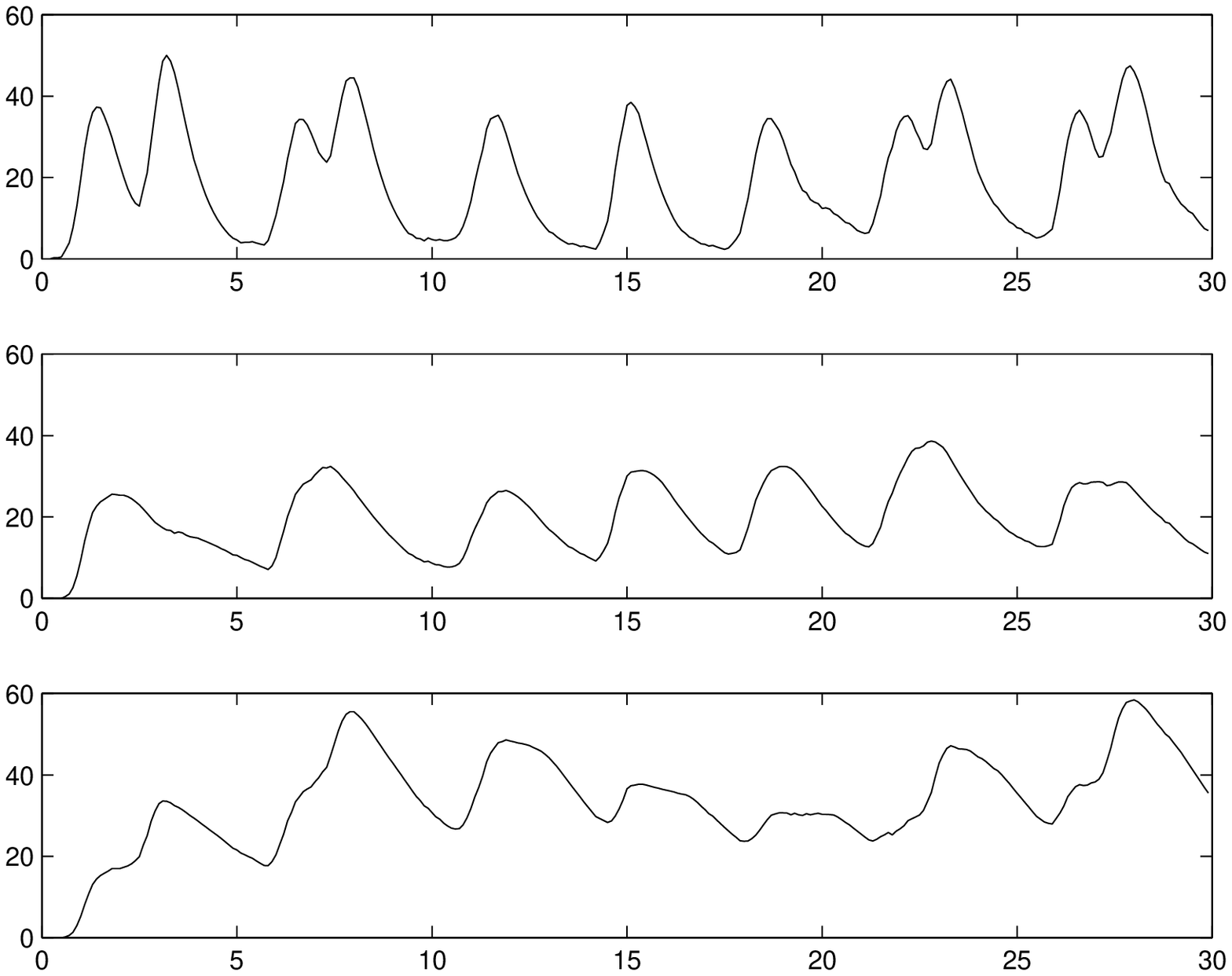}
\caption{\footnotesize Simulating effector signaling with different 
binding times $\theta$=1 (top), $\theta$=2 (middle) $\theta$=3.2 (bottom)
Short binding times separate more signals, longer binding times generate 
broader, lower signals when in phase. Mismatches of binding times with input 
signal frequency (bottom) generate the largest, but least modulated 
integrated signals.}
\label{theta10-20}
\end{center}
\end{figure}

We assume that a receptor can receive separate signals on the order of 
about 100 ms each. Binding times $\theta$ are typically one order of magnitude 
more, e.g. between 1-2 s, and resensitization times are equally variable 
(e.g., 2-4 s, the absolute values depend on the type of 
receptor \cite{KraselCetal2004}).
When a receptor is activated, and a G-protein is associated with the receptor
complex,
it generates a signal proportional in length to its binding time. 
A receptor can produce more signals, when 
binding times are short. 
When binding times are longer, the window-size for integration increases 
(correlation filter), when they are shorter, the window-size decreases 
(decorrelation filter).
We see this reflected in the output vector, comparing shorter 
with longer binding times, everything else being equal
(cf. Fig.~\ref{theta10-20}).
This would allow, for instance, to either combine or separate signals 
mediated by different ligands, such as by monoamine/neuropeptide release 
concurrency.

There may be a specific structure of the input signal: 
periods of high 
coincidence of signals alternating with low incidence of signals.
(cf. Fig.~\ref{theta10-20}, left). If binding times are not well-matched 
to the temporal structure of the input, the output becomes similar to an 
ongoing tonic signal with few modulations (Fig.~\ref{theta10-20}, bottom). 

\vspace{.5cm}
{\large Competition for effectors.}

The main insight into effective competition is the idea that the probability 
of losing an effector protein is inversely correlated to the activity at 
the receptor site ('sticky effectors').

\begin{figure}[htb]
\hspace{0.8cm}
\includegraphics[height=3cm,width=6cm]{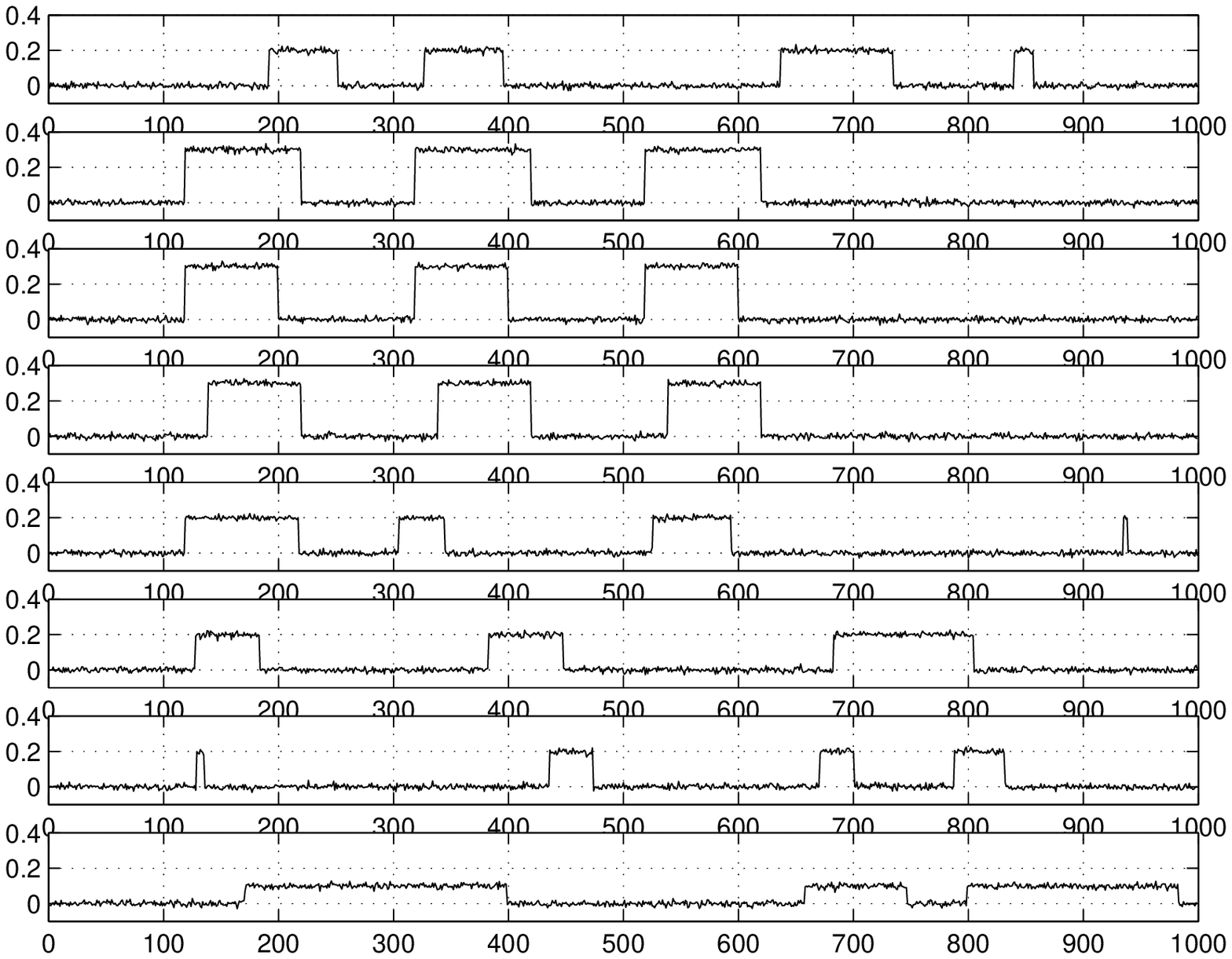}
\hspace{0.8cm}
\includegraphics[height=3cm,width=6cm]{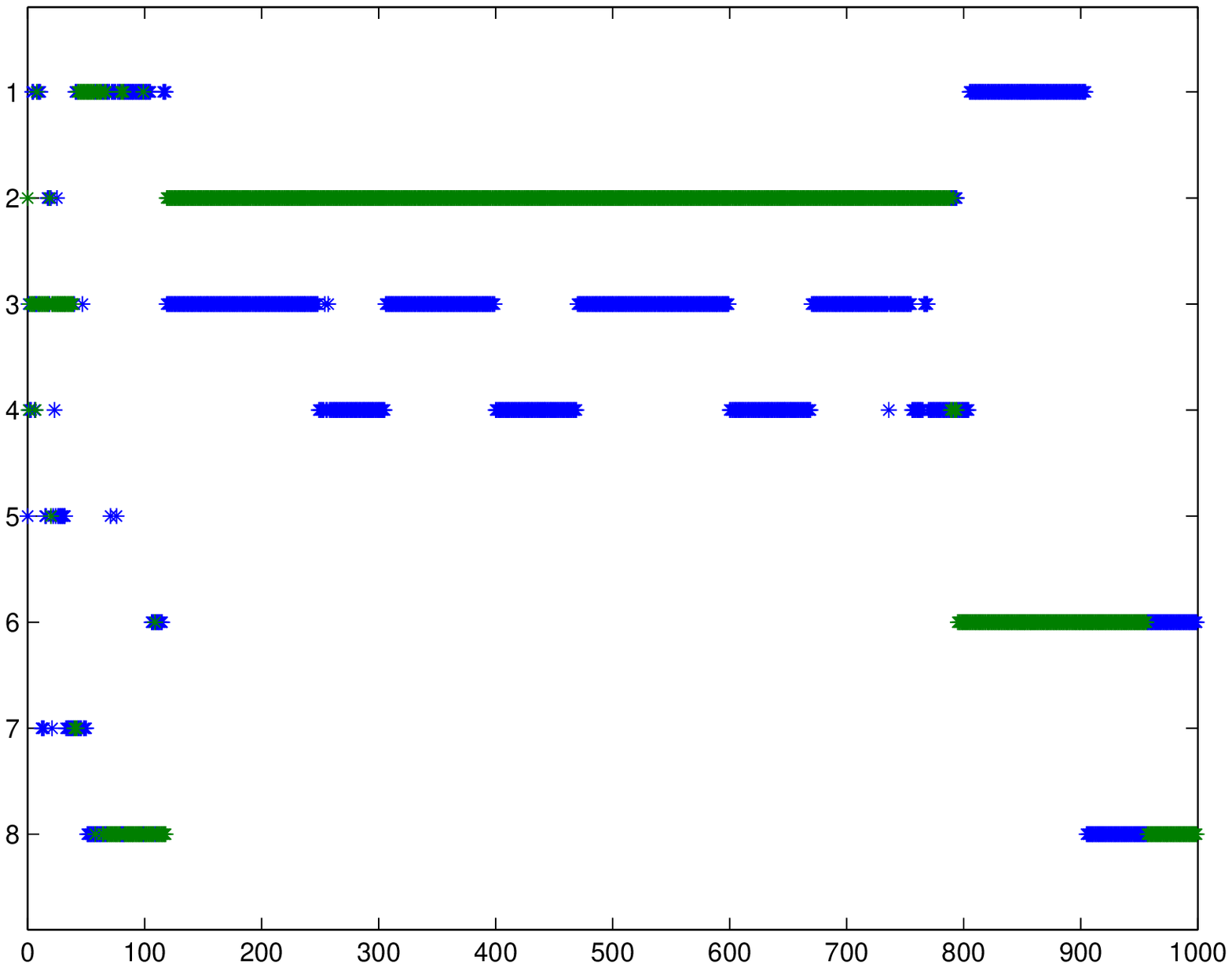}
\caption{\footnotesize (Left) Input signals to 8 different receptors, where signals 2, 3 and 4 
show significant correlation. (Right) Allocation of effectors to input 
units. Receptors 2, 3 and 4 have strongest coupling.}
\label{figure-2}
\end{figure}

This general principle, which can be implemented in a number of different 
ways, guarantees that strong inputs 
have a high probability of continuing influence on the output value and
redistribution of effectors happens only during periods of 
low receptor activation.
We illustrate this mechanism with 
variable input signals for 8 receptors (Fig.~\ref{figure-2},
left). Selection of the most important signals is facilitated by competition 
for a scarce set of effectors. In this case, receptors 2-4 show the 
most effector coupling, while assignment otherwise is probabilistic.

\vspace{.5cm}
{\large Dynamic regulation of effector pool.}

\begin{figure}[htb]
\hspace*{1cm}
\includegraphics[height=1.8cm,width=6cm]{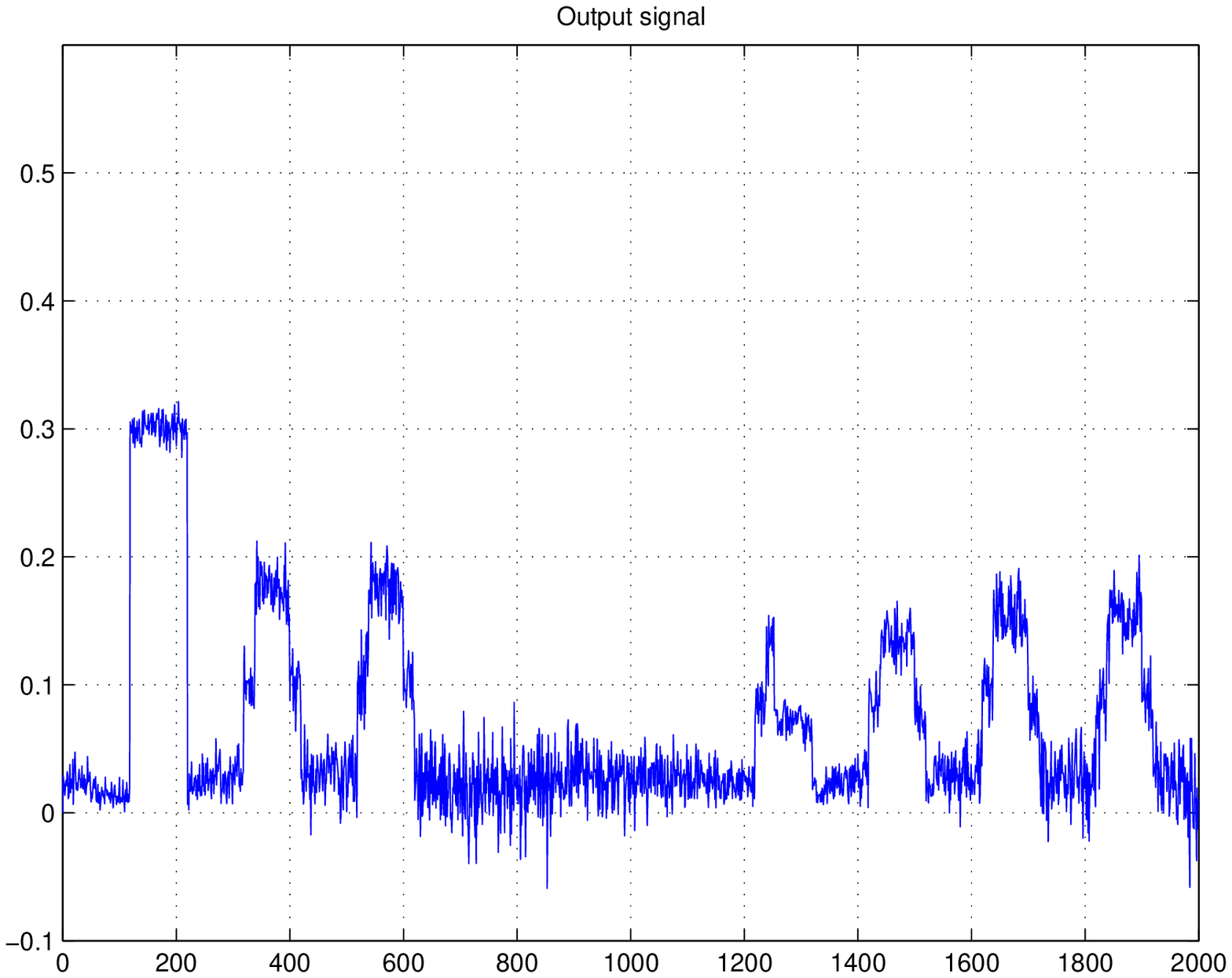}
\hspace*{1cm}
\includegraphics[height=1.8cm,width=6cm]{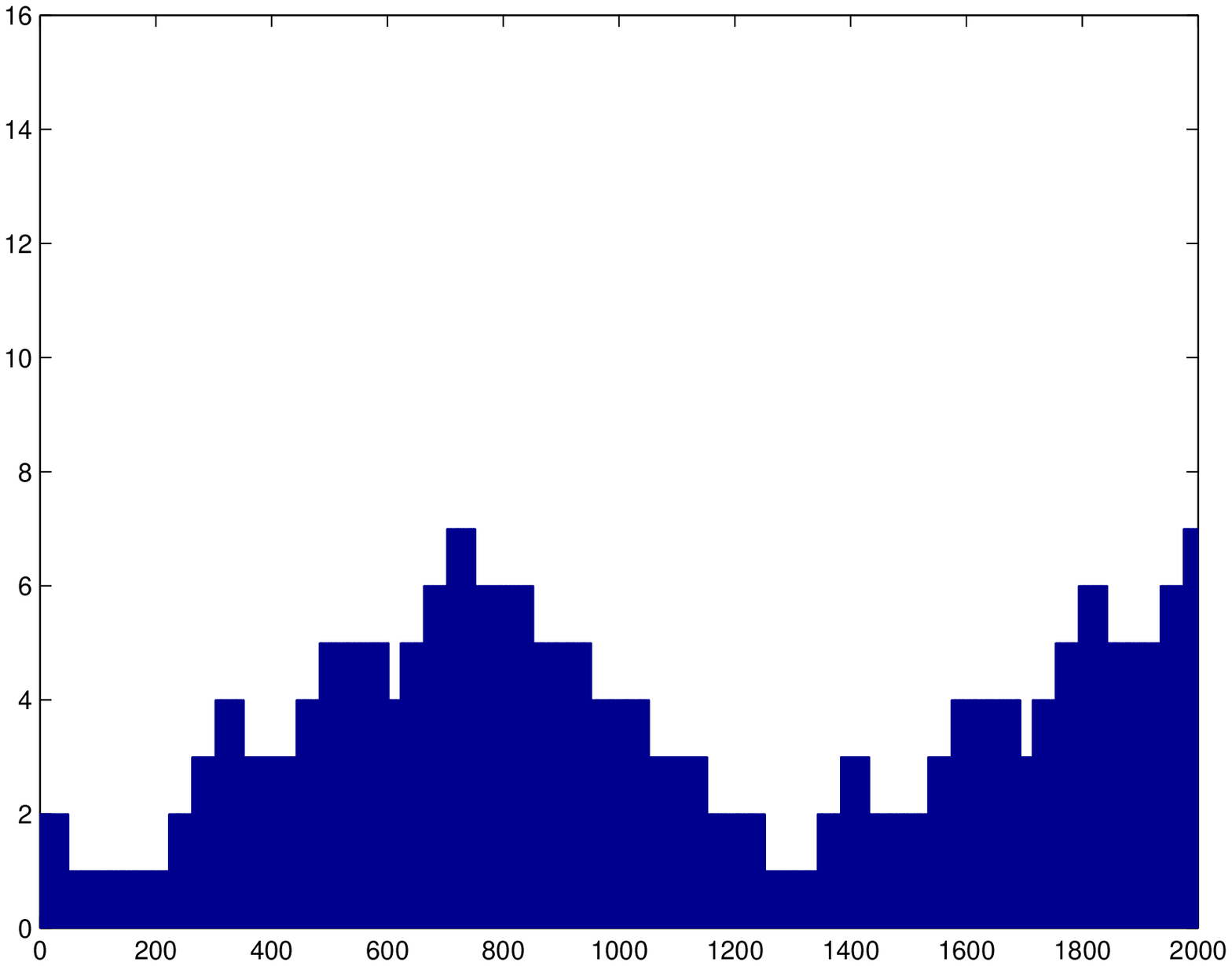}
\caption{\footnotesize The output signal (left) regulates effector pool
size (right). Note the increase of pool size for time units 2-6 and 
decrease for time units 8-14. Individual signals are diminished by 
pronounced temporal integration in regulating pool size.}
\label{figure-3}
\end{figure}

The effector pool size which determines how many input signals are being 
transmitted depends on the output units of the system 
(see Fig.~\ref{figure-1}).
For instance, effector pool size may increase for focused output signals with 
sharp peaks and degrade when signals are weak or diffuse.
With appropriate delays, we will then observe an oscillation between 
periods of focused 
signal transduction (high receptor activity) and search for a strong 
signal (low receptor activity) (Fig.~\ref{figure-3}).
When signal transmission is highly selective, focused output signals are 
generated.  This output signal then increases effector pool size, and thus 
resets the system to a broader tuning in order to subsequently increase its 
chances of picking up new signals. Thus sharp focused peaks in the output 
signal lead to increases of effector size by feedback from the output signal.
Accordingly, effector pool size decreases when signals are broad and 
unspecific.
This may increase the probability of enhancing a new, specific signal through 
competition by receptors for scarce effectors.

\vspace{.5cm}
{\large Conclusion.}
\label{conc}

The interesting result for the role of binding times as cross-correlation 
filters is that intracellular properties are able to either integrate or 
separate 
extracellular signals which occur in close temporal proximity. This is 
a central task for a neuron with a multitude of different G-protein 
coupled receptors. Furthermore, for phasic signals, when information transfer 
is lowest (tonic signals with few modulations), plasticity may be induced 
with high efficiency. This relation between low rates of information transfer 
which induce maximal plasticity requires further attention.
Competition for effectors or binding partners in signal transduction pathways
is a fairly ubiquitous phenomenon in intracellular computation. 
Here the concept of localized efficient computation
maximizes information transfer when high activity events occur.
Adjustment of effector pool size with long delays may induce an oscillation 
of effector availability that coincides with an oscillation of 
focused information transfer by high competition and broad signal search 
by low competition.

\footnotesize 
\def\thebibliography#1{\section*{{\Large\bf \ }
}\list
{[\arabic{enumi}]}{\settowidth\labelwidth{[#1]}\leftmargin\labelwidth
\advance\leftmargin\labelsep
\usecounter{enumi}}
\def\newblock{\hskip .11em plus .33em minus .07em}
\sloppy\clubpenalty4000\widowpenalty4000
\sfcode`\.=1000\relax}
\let\endthebibliography=\endlist
\setlength{\baselineskip}{4.2mm}

\end{document}